# Deep learning piston aberration control of fiber laser phased array by spiral phase modulation

Jing Zuo,[1,3,4,†] Haolong Jia,[2,3,4,†] Chao Geng,[1,3,*] Qiliang Bao,[2,3,**] Feng Li,[1,3] Ziqiang Li,[1,3,4] Jing Jiang,[2,3] Yunxia Xia,[2,3] Fan Zou,[1,3,4] Xu Yang,[1,3,4,5] Xinyang Li,[1,3] and Bincheng Li,[5]

[1] *Key Laboratory on Adaptive Optics, Chinese Academy of Sciences, Chengdu 610209, China*

[2] *Key Laboratory of Optical Engineering, Chinese Academy of Sciences, Chengdu 610209, China*

[3] *Institute of Optics and Electronics, Chinese Academy of Sciences, Chengdu 610209, China*

[4] *University of Chinese Academy of Sciences, Beijing 100049, China*

[5] *University of Electronic Science and Technology of China, Chengdu 611731, China*

*\* blast_4006@126.com*

*\*\* baoqiliang@ioe.ac.cn*

**Abstract:** The stochastic parallel gradient descent (SPGD) algorithm is usually employed as the control strategy for phase-locking in fiber laser phased array systems. However, the convergence speed of the SPGD algorithm will slow down as the number of array elements increases. To improve the control bandwidth, the convolutional neural network is introduced to quickly calculate the initial piston aberration in a single step. In addition, the irrationality of the commonly used Mean Square Error (MSE) evaluation function in existing convolutional neural networks is analyzed. A new evaluation function NPCD (Normalized Phase Cosine Distance) is proposed to improve the accuracy of the neural networks. The results show that the piston aberration residual is 0.005 and the power in the bucket (PIB) is 0.993 after accurate preliminary compensation, which means that the system directly enters the co-phase state. We also demonstrate the robustness and scalability by adding additional disturbance and expanding the scale of the array.

## 1. Introduction

A near Diffraction-Limited equivalent large-aperture output laser can be obtained by correcting the phase-type (piston) aberration in the fiber laser phased array (FLPA), which can be applied to laser transmission, free-space optical communication, Lidar et al [1-5]. The core challenge of achieving a Diffraction-Limited output laser of FLPA is to correct and lock the piston aberration quickly between the sub-beams. In the past, researchers usually use the stochastic parallel gradient descent (SPGD) algorithm to achieve phase locking [13-14]. However, the control bandwidth of the SPGD algorithm is inversely proportional to the $n^{0.5}$ (n is the number of sub-apertures). If some constraint information (such as initial piston aberration) can be obtained in advance, the complicated iteration process of the SPGD algorithm can be avoided, and the real-time performance of the system will be greatly improved. Fortunately, the phase retrieval method based on the diffraction field image is a promising approach to calculate the initial piston aberration [6, 7, 22, 23]. In this case, the FLPA can directly obtain a co-phase state.

With the recent advancement in machine learning, in particular deep learning and reinforcement learning, researchers have recognized that the complex nonlinear relation between far-field image and piston aberration could be mapped by neural network so the piston aberration may be predicted by one-step without iteration optimization. However, this method still faces two main challenges. One is that it is difficult to produce a large enough experimental dataset to train the model. Another is the phase ambiguity. In the FLPA system with a tiled-aperture configuration, by rotating the initial piston aberration 180 degrees and flipping it, the new far-field image will be consistent with the initial one. This is a physical constrains problem just as human beings cannot distinguish the identity of two twins who look the same as each other. For the first challenge, deep learning focuses on feature extraction and requires a large number of data sets to train the neural network to establish a proper nonlinear mapping. Different from deep learning, reinforcement learning could map the nonlinear relationship by the reward mechanism without a large number of pretreatment datasets. In 2021, Henrik Tünnermann and Akira Shirakawa et al derive a suitable control policy without any explicit modeling by using reinforcement learning [22]. In that case, the problem for phase locking is converted to optimize a reward parameter (the beam quality) to obtain a co-phase state. However, as reinforcement learning needs the training time if the state of the system shifts, the robustness and scalability remain questionable in practical experiments. Also in 2021, Shpakovych M et al proposed a quasi-reinforcement learning algorithm to combine the sub-beams with more than hundreds of elements [23]. They segment a plane wave into the array beams and propagate it to the far-field. The piston aberration is only generated by the spatial light modulator (SLM) in the optical path so the change of far-field image is only determined by the output phase of SLM, which avoids the influence of dynamic random aberration. They demonstrated that reinforcement learning is expected to be applied in practical systems and maintain efficient scalability. However, the method may not be suitable for a FLPA system based on master oscillator power amplifier (MOPA) configuration. Only in this configuration, a high brightness combined-beam could be generated in practical applications. While the piston aberrations of each sub-beam are randomly distributed and constantly changing, and the changes of the far-field image are caused by the output phase by SLM as well as the dynamic random aberrations in that case, which will affect the performance of this method. In addition, the solving method of the phase ambiguity has not been discussed in the front two works, which is another question about the robustness of their system.

Recently, with the fast development of deep learning, the convolutional neural network (CNN) presents a superior performance in image recognition and its robustness has greatly improved. A large number of simulation data with labels could be used to train CNN. The well-trained model can predict the piston aberration in the practical system, as long as the simulation model is accurate. In 2019, Hou *et al* used a single-frame defocused plane image as the input data set and the Mean Square Error (MSE) as the evaluation function to coarsely predict the piston aberration [8, 15]. Then, they used the SPGD algorithm for further iteration to achieve a co-phase output of FLPA, according to CNN's output. However, this method still has rooms for improvement: (1) iteration operation is still needed to achieve a convergence state; (2) phase ambiguity (non-uniqueness) may occur if input data set (single-frame defocused image) is not appropriately selected. This non-uniqueness may affect the generalization ability of CNN, and result in worse results after initial compensation as compared to that without compensation; (3) The piston aberration is randomly distributed with [0, 2π], and the MSE cannot correctly evaluate the distance between the output piston of CNN and the ground truth piston.

To address these issues, in this paper we propose using two-frame focused plane images as the input data set for improving retrieval precision. This proposed method is verified with a 7 elements FLPA in which the wavefront phase profile is modulated precisely by using the phase modulator (PM) device in the FLPA so that two different images can be easily obtained in the far-field. With the proposed method, a co-phase state is directly obtained in FLPA and the SPGD algorithm is employed to keep the co-phase state. Due to its generality, the method can be applied to predict other types of aberrations [24].

**2. Principle**

*2.1 Structure of 7 elements FLPA*

The diagram of the FLPA is shown in Fig. 1. The sub-apertures are arranged in a regular hexagon shape. The sub-apertures diameter $d = 28$ mm, the distance between adjacent sub-apertures centers $s = 31$ mm, the beam wavelength $\lambda = 1064$ nm. By controlling the piston aberration between the sub-beams, an equivalent large-aperture near Diffraction-Limited output laser beam can be obtained in the far-field.

The laser source is equally divided into 7 elements by a fiber splitter, the power of each sub-beam is amplified by a power amplifier (PA). After passing through the PA, each sub-beam is connected to a PM, which is employed to correct the phase-type (piston) aberration. Each PM is connected to the collimator array's tail fiber and then the beam is emitted into free space. The output sub-beams are focused by a transform lens with a focal length $f_{lens} = 2$m and then the focused sub-beams are split by a beam-splitter (BS) into two beams. One beam is sent to a $10\times$ micro-objective (MO) and detected by a high-speed CCD camera to obtain the far-field image for phase-locking. Another beam is transformed to the target face for practical application [25].

CNN can quickly retrieve the initial piston aberration, but phase modulation should be added to obtain accurate input data in each retrieval step. Considering that when the system is working, a spiral phase shape modulation cannot continuously be added as it causes the shape of the far-field spot to change continuously that further affect the practical application, CNN is employed to accurately retrieve the piston aberration in the first stage, and then the SPGD algorithm is used to maintain this convergence state. When the evaluation function drops to the non-convergence state, the above operation can be performed again to make the system quickly return to the phase-locking state where the phase-locking state can be judged by the current beam quality (analyzed in section 4).

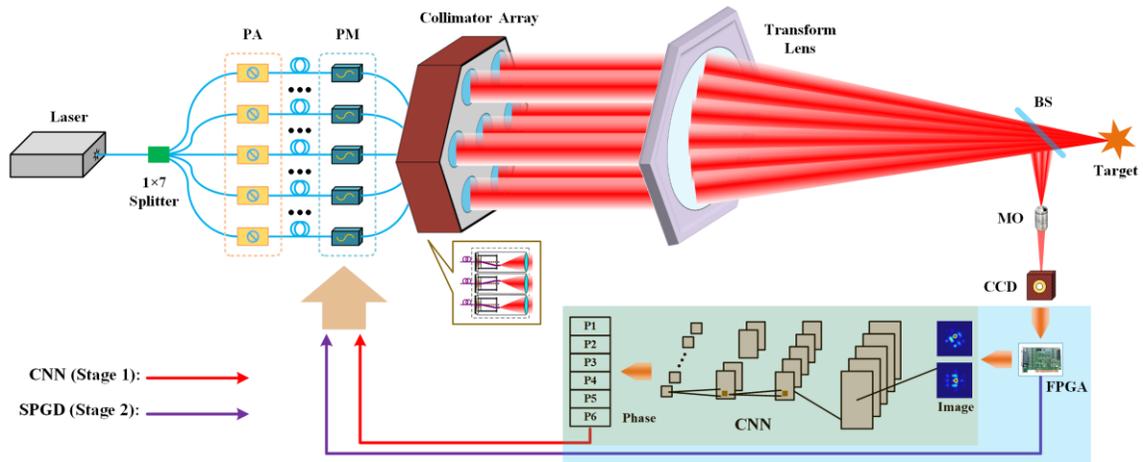

Fig. 1. Schematic diagram of 7 elements FLPA

## 2.2 Multiple-solution issue arising from the rotational conjugate of the initial piston

The FLPA system can be viewed as a multi-beam interference model [20], and the near-field complex amplitude of the nth sub-beam $U_{near_n}$ can be expressed as

$$U_{near_n}(x, y, 0) = \frac{A}{\omega_0} \exp\left[-\frac{(x-x_n)^2 + (y-y_n)^2}{\omega_0^2}\right] \quad (1)$$

where $A$, $\omega_0$, and $(x_n, y_n, 0)$ are the amplitude, waist radius, and central position of the $n_{th}$ sub-beam in the near-field. The parameters are as follows: $A = 1$, $\omega_0 = 11$ mm.

$$E_{near} = \exp[i \cdot \Psi(x, y)] \cdot \sum_{n=1}^{7} U_{near_n} \quad (2)$$

$E_{near}$ represents the superposed Gaussian sub-beams modulated by a random phase screen with a random range of 0~2π rad in the corresponding sub-aperture. When the amplitudes of sub-beams are equal, we only consider the influence of piston aberration. In Fourier optics, the relationship between the complex amplitude of far-field $E_{far}$ and near-field $E_{near}$ can be expressed as

$$E_{far}(x_0, y_0) = \iint E_{near}(x, y) \exp[-i2\pi(ux + vy)] dx dy \quad (3)$$

Where $(x, y)$ and $(x_0, y_0)$ are the rectangular coordinates on the near-field plane and the far-field plane, respectively, $u$ and $v$ are spatial frequencies in which $u = \frac{x_0}{\lambda f_{lens}}, v = \frac{y_0}{\lambda f_{lens}}$. According to the Euler formula, $E_{near} = \cos\Psi(x, y) + i \cdot \sin\Psi(x, y)$. By rotating the initial piston aberration $\Psi(x, y)$ 180 degrees and then flipping it, we can obtain a new piston aberration $\Psi' = -\Psi(-x, -y)$ and its far-field $E'_{far}$. According to the Euler formula, $E_{far}$ and $E'_{far}$ can be newly expressed as

$$\begin{aligned} E_{far}(x_0, y_0) &= \iint [\cos\Psi(x, y) + i \cdot \sin\Psi(x, y)] \cdot [\cos 2\pi(ux + vy) - i \cdot \sin 2\pi(ux + vy)] dx dy \\ &= \iint \{[\cos\Psi(x, y) \cdot \cos 2\pi(ux + vy) + \sin\Psi(x, y) \cdot \sin 2\pi(ux + vy)] \\ &\quad + i \cdot [-\cos\Psi(x, y) \cdot \sin 2\pi(ux + vy) + \sin\Psi(x, y) \cdot \cos 2\pi(ux + vy)]\} dx dy \\ &= \iint \{\cos[\Psi(x, y) - 2\pi(ux + vy)] + i \cdot \sin[\Psi(x, y) - 2\pi(ux + vy)]\} dx dy \end{aligned} \quad (4)$$

$$\begin{aligned}
E'_{far}(x_0, y_0) &= \iint \{\cos[-\Psi(-x,-y)] + i \cdot \sin[-\Psi(-x,-y)]\} \cdot [\cos 2\pi(ux+vy) - i \cdot \sin 2\pi(ux+vy)]dxdy \\
&= \iint [\cos \Psi(-x,-y) - i \cdot \sin \Psi(-x,-y)] \\
&\quad \cdot (\cos\{-2\pi[u(-x)+v(-y)]\} - i \cdot \sin\{-2\pi[u(-x)+v(-y)]\})d(-x)d(-y) \\
&= \iint \{\cos \Psi(-x,-y) \cdot \cos 2\pi[u(-x)+v(-y)] + \sin \Psi(-x,-y) \cdot \sin 2\pi[u(-x)+v(-y)]\} \\
&\quad + i \cdot [\cos \Psi(-x,-y) \cdot \sin 2\pi[u(-x)+v(-y)] - \sin \Psi(-x,-y) \cdot \cos 2\pi[u(-x)+v(-y)]d(-x)d(-y) \\
&= \iint \cos\{\Psi(-x,-y) - 2\pi[u(-x)+v(-y)]\} - i \cdot \sin\{\Psi(-x,-y) - 2\pi[u(-x)+v(-y)]\}d(-x)d(-y)
\end{aligned}$$

(5)

$E_{far}$ and $E'_{far}$ have the same real components and opposite imaginary components. In Fourier optics, the far-field intensity distribution is equivalent to the squared modular operation of the complex amplitude

$$|E_{far}|^2 = |E'_{far}|^2 \tag{6}$$

This means that $E_{far}$ and $E'_{far}$ correspond to the same far-field image. In supervised learning, equation (6) indicates that a single input (far-field image) corresponds to multiple labels (piston aberration). As shown in Figure 2, the mapping relationship is morbid, so that the supervised learning in CNN cannot decipher whether the label is true or not.

As shown in Fig. 3, we superimpose a spiral phase shape on the tangential direction of the sub-beams, which has a radians range of equal intervals from 0 to $2\pi$ on $\Phi(x, y)$. This spiral phase shape can be obtained by controlling the PM without adding any additional optics so the integration level of FLPA is greatly improved, which can solve the non-uniqueness problem. In fact, the shape of the modulation phase can be any asymmetric shape. However, by modulating the expected piston aberration of the sub-beams with a shape of spiral phase, the orbital angular momentum beam can be easily obtained, which has unique applications in many fields [15]. By using spiral phase shape modulation, the modulated complex amplitude $E_{newfar}$ and $E'_{newfar}$ can be expressed as

$$\begin{aligned}
E_{newfar}(x_0, y_0) &= \iint \{\cos[\Psi(x,y) + \Phi(x,y)] + i \cdot \sin[\Psi(x,y) + \Phi(x,y)]\} \\
&\quad \cdot [\cos 2\pi(ux+vy) - i \cdot \sin 2\pi(ux+vy)]dxdy \\
&= \iint \{\cos[\Psi(x,y) + \Phi(x,y) - 2\pi(ux+vy)] \\
&\quad + i \cdot \sin[\Psi(x,y) + \Phi(x,y) - 2\pi(ux+vy)]\}dxdy
\end{aligned} \tag{7}$$

$$\begin{aligned}
E'_{newfar}(x_0, y_0) &= \iint \{\cos[-\Psi(-x,-y) + \Phi(x,y)] + i \cdot \sin[-\Psi(-x,-y) + \Phi(x,y)]\} \\
&\quad \cdot [\cos 2\pi(ux+vy) - i \cdot \sin 2\pi(ux+vy)]dxdy \\
&= \iint \cos\{\Psi(-x,-y) - \Phi(x,y) - 2\pi[u(-x)+v(-y)]\} \\
&\quad - i \cdot \sin\{\Psi(-x,-y) - \Phi(x,y) - 2\pi[u(-x)+v(-y)]\}d(-x)d(-y)
\end{aligned} \tag{8}$$

From Eqs. (7) and (8), we have $|E_{newfar}|^2 \neq |E'_{newfar}|^2$ when $\Psi(x,y) + \Phi(x,y) \neq \Psi(x,y) - \Phi(-x,-y)$, which means that $E_{far}$ and $E_{newfar}$ can determine a sole initial piston aberration label. In this way, we can solve the non-uniqueness problem as shown in Figs. 3 (b) and (c). Two pairs of far-field images (fig. 2(b)&fig.3(b) and fig.2(d)&fig.3(c)) will correspond to two different piston aberrations.

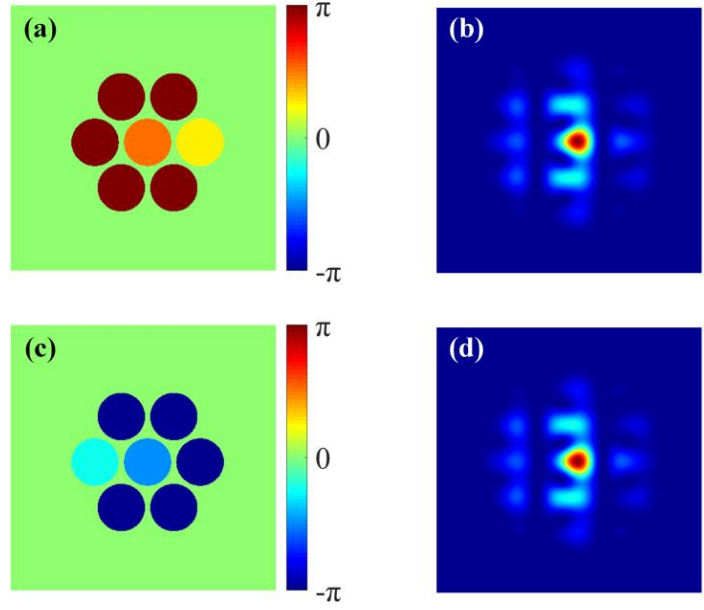

Fig. 2. (a), (b) Initial piston aberration distribution and corresponding far-field image

(c), (d) Piston aberration distribution and corresponding far-field image after rotating and conjugating

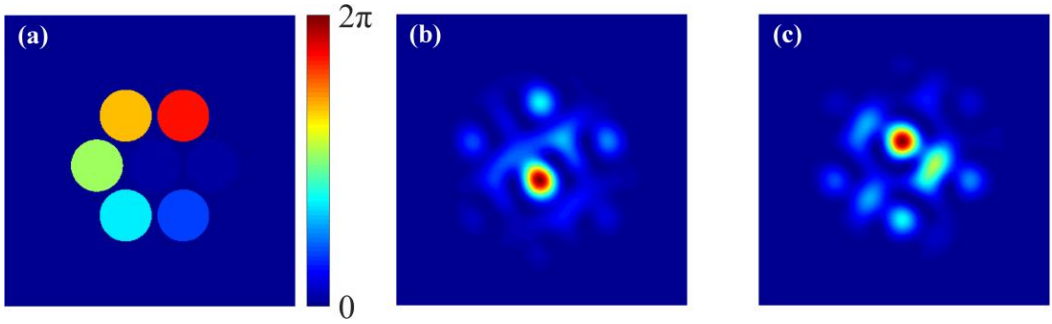

Fig. 3. (a) Spiral phase shape produced by PM

(b), (c) Different spiral phase shape modulation produces different far-field images

## 2.3 Multi-solution issue caused by a redundant piston

Since there is a piston tolerance $\Delta\varphi$ between each sub-beam, such a redundant piston will make $\boldsymbol{\varphi}_{\mathbf{in}} = \left[ \varphi_{in}^1 \cdots \varphi_{in}^7 \right]$ and $\boldsymbol{\varphi}_{\mathbf{in}} = \left[ \left( \varphi_{in}^1 + \Delta\varphi \right) \cdots \left( \varphi_{in}^7 + \Delta\varphi \right) \right]$ have the same far-field image, results in non-uniqueness issue for network optimization. Our solution is to take the piston aberration $\varphi_{in}^1$ as reference and subtract this aberration value from the other six piston aberrations to obtain a relative piston distribution set $\boldsymbol{\varphi}_{\mathbf{in}} = \left[ 0 \cdots \varphi_{in}^7 - \varphi_{in}^1 \right]$. Via this operation, the pathological problem is overcome.

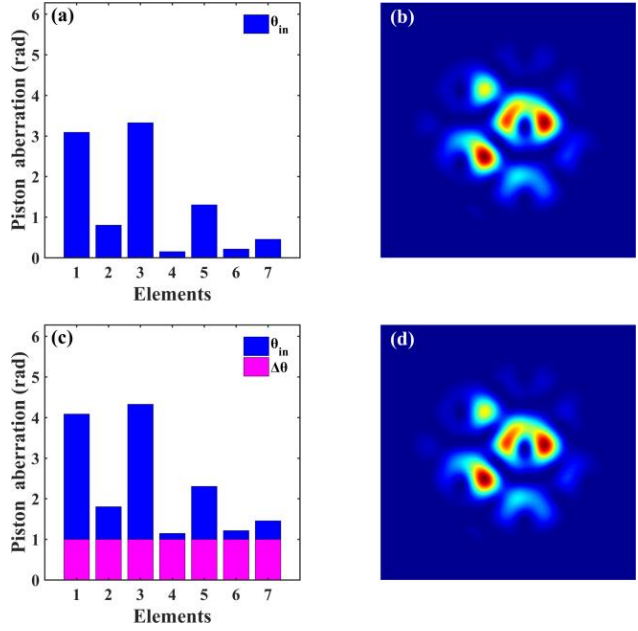

Fig. 4. Far-field images corresponding to a set of redundant pistons

## 3. Convolutional neural network

### 3.1 Neural network structure

The convolutional neural network in our work is based on the residual network module [16], as shown in Fig. 5. The structure in Fig. 5(a) outputs feature maps with the same size as the inputs after feature extraction. The structure in Fig. 5(b) can extract feature from inputs, and then down-sample them. The shortcut connections of residual network can spread the feature extracted from shallow layers to deep layers, so that they enhance the expression ability of neural network by feature reuse. The connections could avoid the risk of over-fitting at the same time. A 3×3 convolution kernel is used in the main path of the module. In the down-sampling module, the second 3×3 convolution kernel is replaced by a 2×2 convolution kernel whose stride is 2. In shortcut connections, identity mapping is used when the element number of input feature maps is equal to the output's, otherwise the input is converted to output using a 1×1 convolution. We use the scale factor of 0.5, bilinear sampling in the down-sampling module. We also use batch normalization in our networks to make the training more stable [17]. We choose Mish as the activation function instead of ReLU, because the former shows better performance in stabilizing training and improving accuracy [18-19]. The expression of Mish activation function is as follows

$$f(x) = x \cdot \tanh(\mathrm{softplus}(x)) \quad , \tag{9}$$

where $\mathrm{softplus}(x) = \log(1+\exp(x))$.

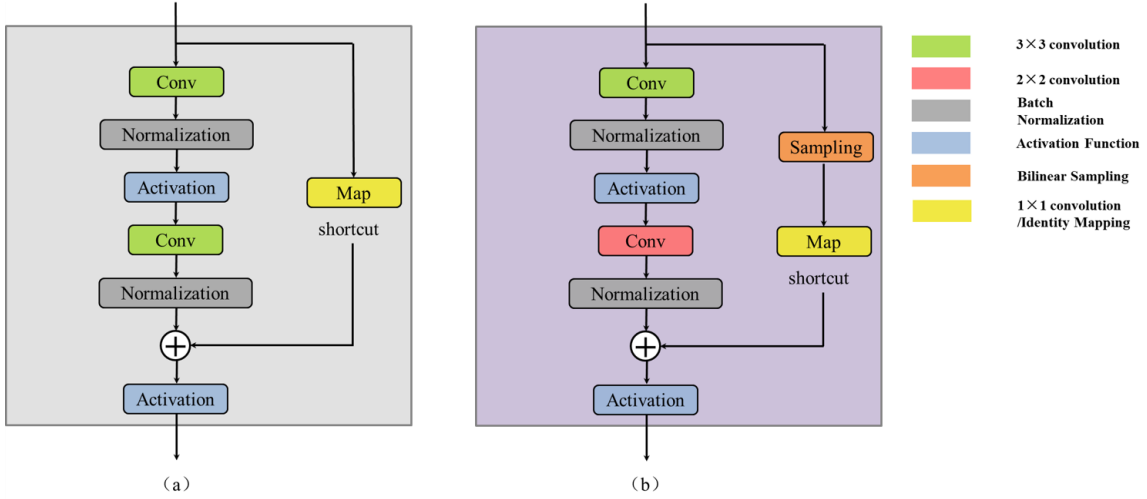

Fig. 5. (a) Residual module (b) Residual module using down-sampling.

(The structure corresponding to different colors is shown in the upper right corner of the picture)

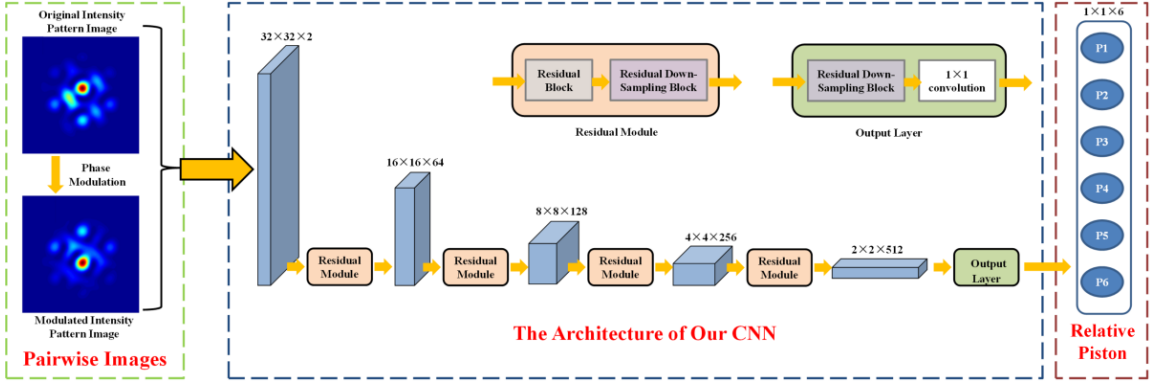

Fig. 6. Framework of our networks

Figure 6 shows the convolutional neural network framework we use. The network consists of 4 residual modules and 1 output layer. Firstly, we take a pair of images containing the original far-field image and the modulated one as input. Then we resize the images to 32×32 and apply concatenation to them. The elements of 6, size of 1×1, output calculated by our networks, which corresponds to *j*-1 relative piston respectively, where *j*= 7. We map the value of output to [0, 2π] by the following formula

$$f(x) = (\tanh(x)+1) \cdot \pi \quad (10)$$

### 3.2 Evaluation function of CNN

By evaluating the piston residual error, we can compare the performance of CNN with different parameters and determine when to terminate our training. To this end, we need to measure the distance between the network output $\boldsymbol{\varphi}_{out} = \left[\varphi_{out}^{1} \cdots \varphi_{out}^{j-1}\right]$ and the ground truth $\boldsymbol{\varphi}_{gt} = \left[\varphi_{gt}^{1} \cdots \varphi_{gt}^{j-1}\right]$. In previous work, MSE is used to calculate the distance, which is defined as

$$\text{MSE} = \frac{1}{j-1} \sum_{n=1}^{j-1} \left( \varphi_{out}^n - \varphi_{gt}^n \right)^2 \tag{11}$$

However, selecting MSE as evaluation function is not always appropriate. Table 1 shows the ground truth data set and 2 different predictions. As proven in section 2.2 that there is a one-to-one mapping between relative phase and a pair of images (far-field image and modulated image). the proximity of different phases can be judged by the similarity between the corresponding image pairs. Figures 7 (a), (b) and (c) are image pairs corresponding to ground truth, output1 and output2, respectively. Figure 7 (c) is more similar to (a) than (b). In other words, Output2 is closer to the ground truth piston aberration than Output1.

However, from Table 1 MSE between Output1 and ground truth is smaller than that between Output2 and ground truth. This indicates that it is unreliable to use MSE to evaluate the piston residual. The reason is that the piston aberration changes periodically, but MSE cannot reflect this change effectively. For example, $\varphi_1 = \theta_1$ and $\varphi_2 = 2\pi - \theta_2$ are different piston aberration types. They are very close when $\theta_1$ and $\theta_2$ approach 0. However, the distance reflected by MSE reaches a large value. It is obviously contrary to that facts. This unreasonable situation makes MSE unsuitable for evaluation function. Considering the features of piston aberration, we propose Normalized Phase Cosine Distance (NPCD) to evaluate the performance of networks. NPCD is defined as

$$NPCD = \frac{1}{2} - \frac{1}{2(j-1)} \sum_{n=1}^{j-1} \cos\left( \varphi_{out}^n - \varphi_{gt}^n \right) \tag{12}$$

where $n$ represents the index of relative piston aberration $\varphi_{out}^n$, which represents the $n_{th}$ relative piston aberration predicted by our networks, and $\varphi_{gt}^n$ is ground truth of the $n_{th}$ relative piston aberration. A smaller NPCD means a closer prediction of the piston aberration by networks to ground truth. From this formula, it can be seen that when the difference between $\varphi_{out}^n$ and $\varphi_{gt}^n$ is an integral multiple of $2\pi$, NPCD takes the minimum value of 0. When the difference between $\varphi_{out}^n$ and $\varphi_{gt}^n$ is an integral multiple of $\pi$, NPCD takes the maximum value of 1. Besides, the fact that $\varphi_{out}^n$ is greater than $\varphi_{gt}^n$ by $\Delta\varphi$ is equal to that $\varphi_{gt}^n$ is greater than $\varphi_{out}^n$ by $2\pi - \Delta\varphi$, which is also reflected in NPCD.

It can be seen from Table 1 and Fig. 7 that NPCD can correctly reflect the periodic characteristics of piston aberration. Through the distance between the two groups of output in Table 1 and ground truth by using NPCD as evaluation function, we can find that Output2 is closer to Ground Truth, indicating that NPCD is more reliable than MSE in evaluating network output.

Table 1: MSE and NPCD between outputs predicted by 2 networks (under different parameters and ground truth)

| | Relative piston aberration | MSE | NPCD | PIB |
|---|---|---|---|---|
| Ground Truth | (3.178, 6.070, 2.464, 1.711, 4.671, 5.347) | \ | \ | 0.372 |
| Output1 | (3.088, 3.194, 3.253, 3.016, 3.189, 3.162) | 2.929 | 0.457 | 0.274 |
| Output2 | (3.090, 0.000, 2.481, 1.473, 4.701, 5.333) | 6.152 | 0.004 | 0.992 |

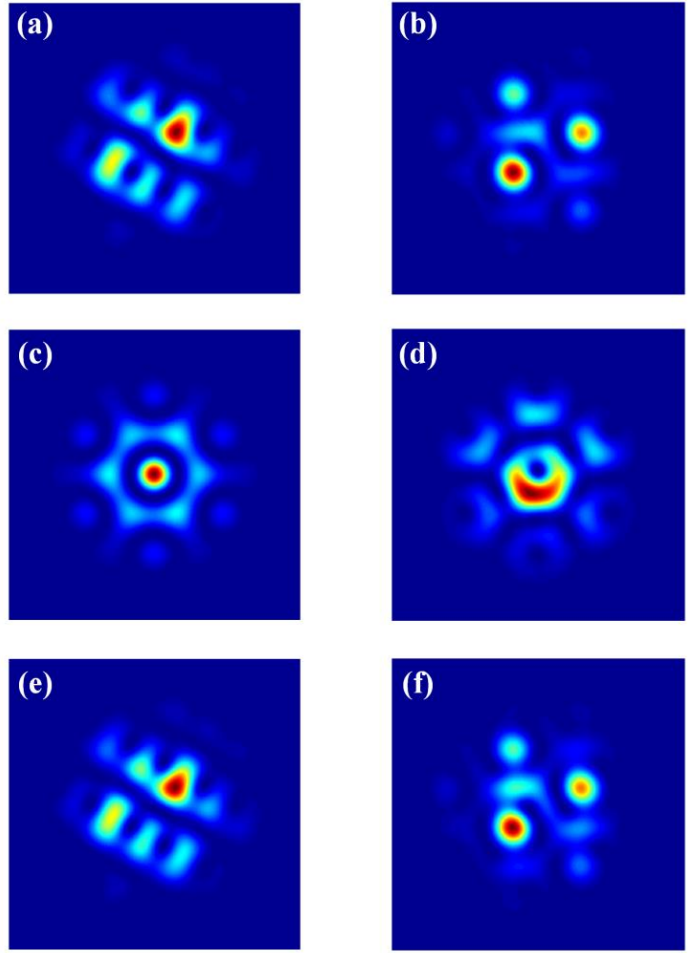

Fig. 7. Initial images (left) and modulated images (right) in the far-field

(a)(b) ground truth, (c)(d) output1, and (e)(f) output2

### 3.3 Training of CNN

Based on section 2, we used MATLAB to generate 36000 groups of 7 elementss far-field images. Each group of images contains an initial image and a modulated image. Each group of images corresponds to a relative initial piston list $\boldsymbol{\varphi_{gt}} = \left[ \varphi_{gt}^2 \cdots \varphi_{gt}^j \right]$. NPCD, which performs better in evaluating the network's effect, is employed in the training.

We used 30000 groups of far-field images and their corresponding piston labels for training, and the remaining 6000 groups for testing. (We map the piston range to [0, 2π], and ground truth piston in the following paragraphs are all from 0 to 2π.) The details in training are as follows: the training steps is 260;

the batch size is 60; we use the above three loss functions to train our networks; we use Adam as optimizer to train network with learning rate $1\times10^{-3}$; we also use dropout and other techniques to avoid over-fitting in training. The server used in the experiment is configured as Intel Core i9-7920X 2.90GHz, NVIDIA GeForce RTX 2080ti. Our models are trained with a RTX 2080 Ti GPU. It takes about 3.6 ms for the trained network to perform a forward calculation. The algorithm can further improve the speed with higher hardware configuration.

## 4. Performance and Disscusion

### 4.1 Performance

The power in the bucket (PIB) is used to evaluate the beam quality in the far-field, which is defined as the power in the Airy disc $(d_{bucket}=2.44\lambda/D)$ of the equivalent aperture divided by the total emission power. Where $D$ = 90 mm, which is the diameter of the circumscribed circle of the emissive array. The value of PIB is influenced by the parameters of FLPA as well as the bucket size. For the FLPA system in this manuscript, the theoretical PIB is $\sigma$ = 0.525. To facilitate data analysis, we have done a normalization process. All the PIBs obtained are divided by $\sigma$ to normalize them to between 0 and 1.

$$PIB = \frac{\iint \text{circ}\,(d_{bucket})\,|E_{far}|^2\,dx_0 dy_0}{\sigma \cdot \iint |E_{far}|^2\,dx_0 dy_0} \tag{13}$$

We use MSE as loss function to train our networks. The details in training are described in Section 3.3. The change curve of mean values of evaluation function varied with training steps is shown in Fig. 8. We test every 10 training steps. The evaluation function we used to describe the piston residual is NPCD (Eq. 12). After 260 epochs of training, the piston residual can be reduced to 0.005. The lower the value is, the closer the network output value is to the ground truth value, and the more accurate the network output is. The curve shows that the value of NPCD remains relatively stable in the later stage of training, and there is no obvious downward trend, which means that our network is well trained and there is no under-fitting.

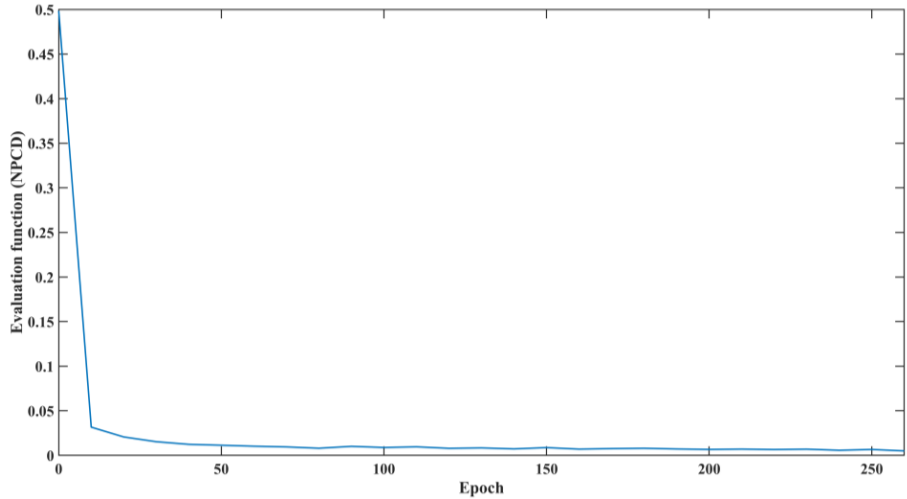

Fig. 8. Curve of evaluation function (NPCD) during training process

Figure 9(b) is the PIB test data scatter diagram calculated from the far-field images that we obtain by compensating the original piston with the output of our networks trained with MSE. Comparing to Fig. 9(a), which is the PIB scatter diagram corresponding to the initial piston aberration, we can find that the PIB's mean value has improved from 0.463 to 0.993 after compensated by the output of CNN. Figure 10(b) shows that there are 5853 groups of data in the range of normalized PIB > 0.9, which suggests that our networks have good estimation performance on most of the data. Figure 11 is the result of the piston aberration compensation with our networks. We can see that the piston aberration predicted by our networks can be fitted with the ground truth (initial piston) with high precision. The compensated far-field images are close to the ideal ones.

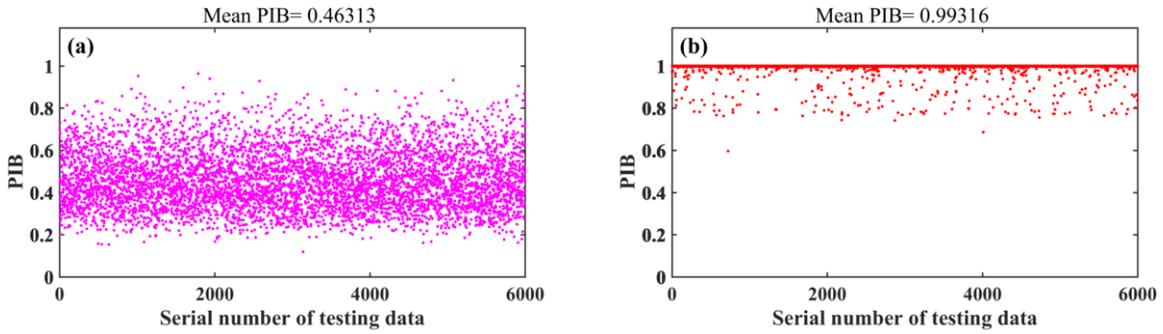

Fig. 9. PIB distribution of 6000 test data

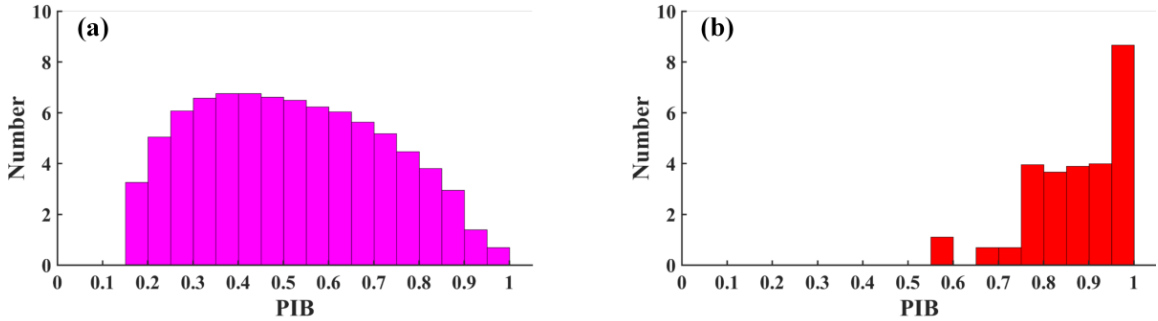

Fig. 10. PIB histogram of 6000 test data

The results show that CNN has satisfactory compensation ability with high precision and the highest average PIB. Our PIB is also higher than 0.97 of Hou *et al* [8, 15]. Moreover, for the 7 elements array, their method saved 30 convergence steps compared with traditional SPGD algorithm. However, our method only needed one step to achieve this state.

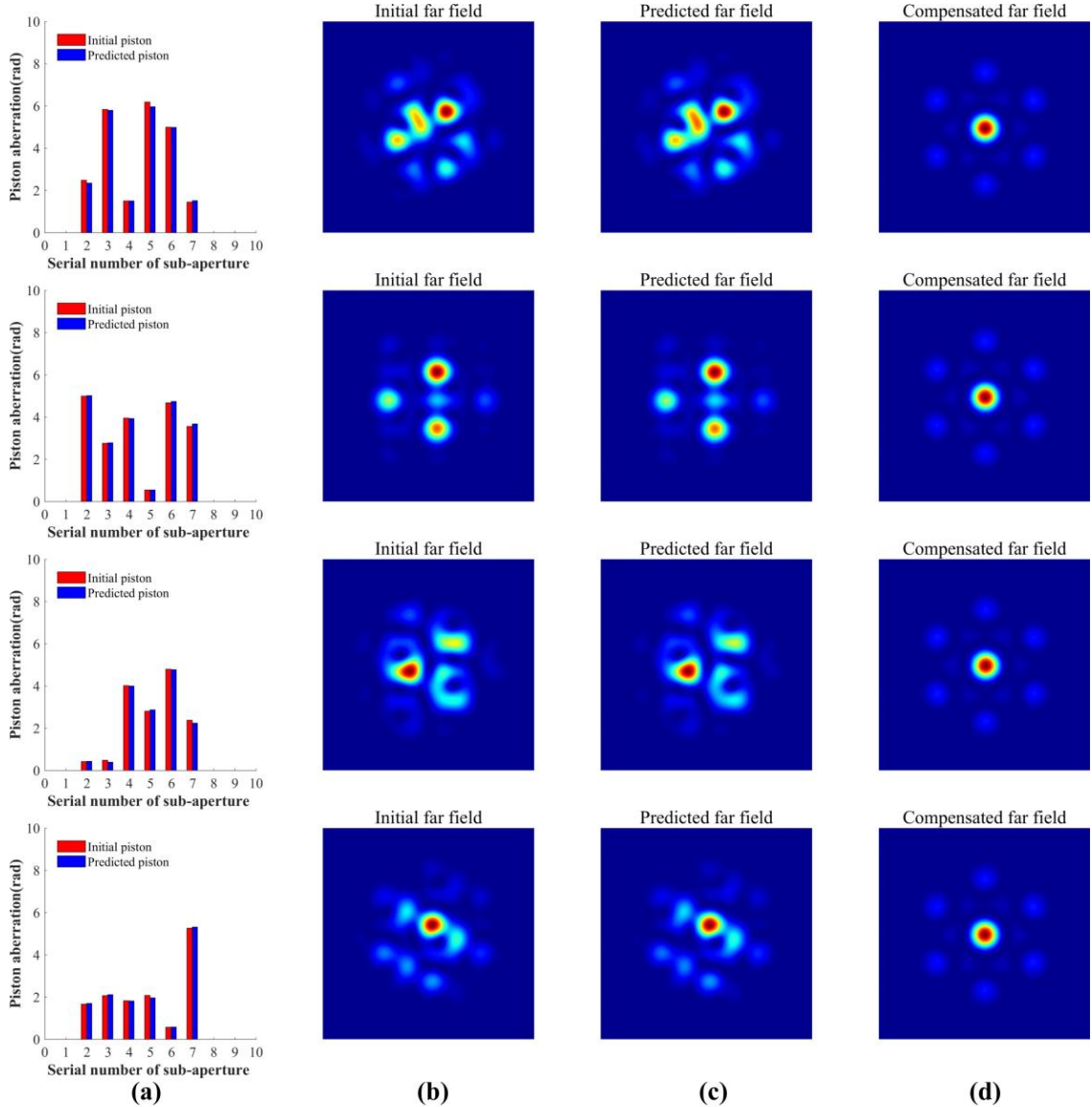

Fig. 11. (a) Comparison of initial piston and predicted piston (b) Far-field images of initial piston
(c) Far-field images of predicted piston (d) Compensated far-field images

*4.2 Compare with SPGD algorithm*

The foregoing content has proved that CNN can achieve fast phase retrieval in 7 elements FLPA system. To further demonstrate the advantages of this method, we use CNN and the traditional SPGD algorithm to compensate the phase under different piston aberrations according to the distribution of the initial PIB range. The comparison between the effects of the two methods is shown in Fig. 12. Each sub-fig contains 30 cases of independent random distribution of initial piston aberrations. We found that the lower initial PIB needs more steps to retrieve piston aberrations by using SPGD. In order to achieve the same accuracy as CNN, SPGD requires 56, 54, 51 and 38 iteration steps for different ranges of initial PIB. Unlike SPGD, CNN has a better retrieve effect for different initial PIB, and a higher PIB output can always be obtained after one step. Table 2 lists the advantages of our methods as compared to the traditional SPGD algorithm.

According to Fig. 12 and table 2, our method is more robust to different input ranges, allowing higher combining bandwidth. In our method, CNN can be regarded as a wavefront sensor, and the piston aberration can be directly predicted without additional iterative optimization. The control bandwidth is only influenced by the performance of the CNN. By designing a reasonable structure, the performance of the CNN can be effectively improved. We only need to fine-tune the parameters of the output layer to fit the number of sub-beams when the number increases, which results in a negligible real-time penalty. Therefore, the control bandwidth of this method can be regarded as unchanged when the number of FLPA elements increases.

CNN can analyze the variety of far-field spot image distortion caused by piston aberration, and then retrieve and compensate the piston aberration in one step. Considering that when the system is working, keeping adding a spiral phase shape modulation would cause the shape of the far-field spot to change continuously that further affect the practical application. When the system converges, the gradient-change of PIB is caused by the dynamic random aberration where the variation of aberration is very small in a short time. SPGD algorithm can recapture the convergence state with a few iterative operations where the time of a single step of SPGD algorithm is 0.48 ms. Thus, we suggest that CNN can be used to accurately retrieve piston aberrations in the first stage, and then the SPGD algorithm can be used to maintain this convergence state.

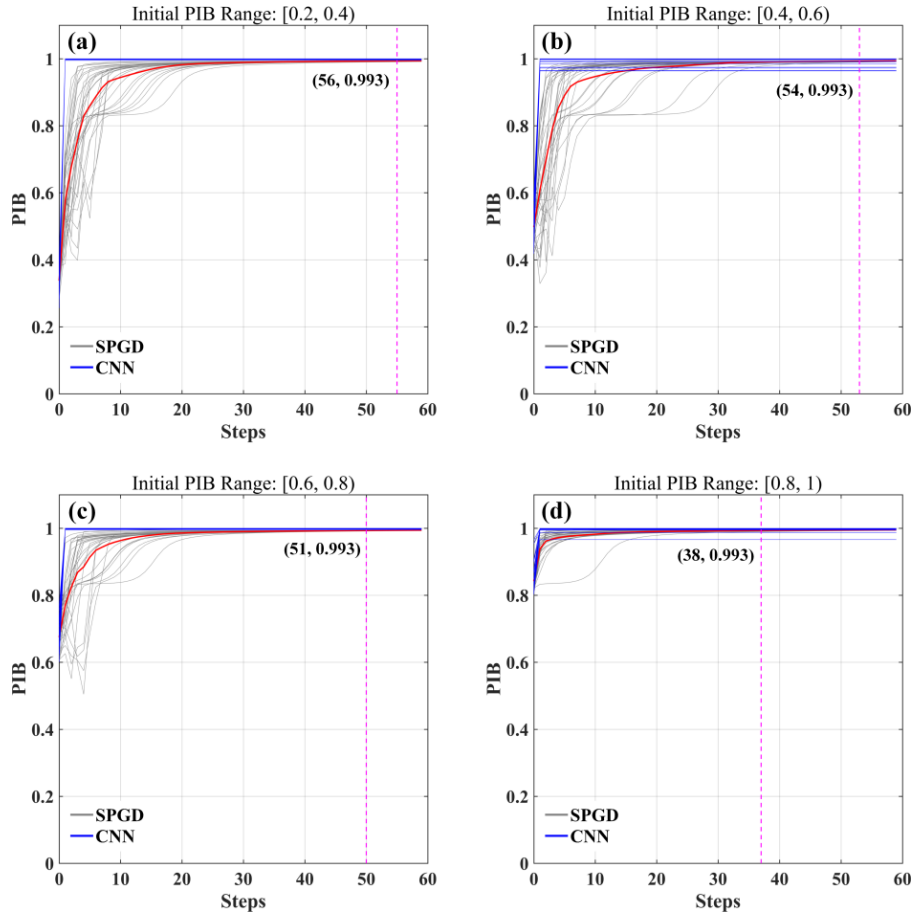

Fig. 12. Comparison of correction effect between CNN (30 cases of blue curve) and SPGD (30 cases of gray curve, and a red curve of average PIB) in different initial PIB ranges

Table 2 Comparison of Deep learning and SPGD in n elements FLPA

| Methods | Deep learning | SPGD algorithm |
|---|---|---|
| Feed back | Image | Image or PIB |
| Control bandwidth | Always 1 step | $\propto n^{0.5}$ |
| Computational complexity | High | Low |
| Detector sensitivity requirement | Low | High |

*4.3 Discussion of stability*

In practical experiments, we may encounter noise from different sources. These noises usually lead to the distribution difference between experimental data and training data, which will affect the performance of the model. In order to evaluate the stability of our CNN, we generated the far-field images with different noises according to the actual situation by simulation and tested our model on these data.

In a practical FLPA system, the PM is controlled by voltage, and there may exist some weak disturbances. An ideal model should be robust to these disturbances. On the contrary, the value of prediction will fluctuate greatly due to weak disturbance in some unstable methods (such as over-fitting models or algorithms facing multi-solution problems), resulting in inaccurate predictions. We selected 200 groups of phase and added random disturbance with amplitude lower than 0.2 to them to simulate voltage noise. As shown in figure 13, we tested our model on the data with phase disturbance. Compensating the initial phase with our predictions will significantly improve the PIB of the combined beam. The average PIB is 0.990, which is close to that on test data. The results show that the phase disturbance has no obvious impact on the performance of our model.

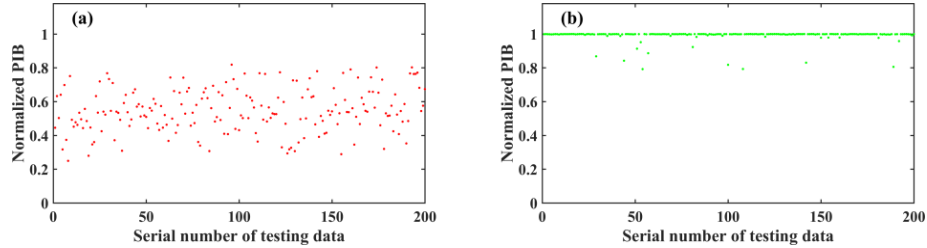

Fig. 13. PIB distribution of 200 test data with random disturbance on PM

(a) Initial PIB distributions, (b) PIB distributions after being compensated by CNN

In the process of image acquisition, noise may generate on the input image due to the instability of the CCD. This kind of noise belongs to Gaussian noise. We simulate this situation by adding Gaussian intensity noise to the initial far-field images in the simulation experiment. As shown in figure 14, the average PIB is 0.992 after being compensated. The result shows the impact on the performance of our model is negligible, which indicates our model is robust to the noise of CCD.

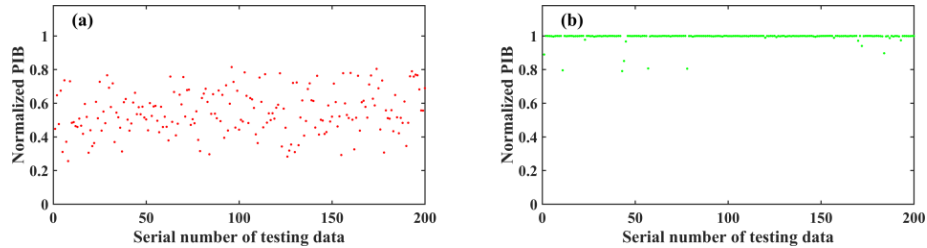

Fig. 14. PIB distribution of 200 test data with Gaussian intensity noise

(a) Initial PIB distributions, (b) PIB distributions after being compensated by CNN

*4.4 Discussion of scalability*

To demonstrate the scalability, our method is extended to a 19-elements PLPA system. All the simulation parameters are equal to that of the real FLPA reported in our previous work [25], as shown in Fig. 15. We retrained and tested our model on the generated data. The test result is shown in Fig. 16, it takes about 3.73 ms for the 19 elements FLPA system to retrieve the piston aberrations where the average PIB is as higher as 0.968. It means that the control bandwidth and precision are hardly affected by the increase of the array elements. Note that the performance of our model on the 19 elements system declines compared with that in the 7 elements system, which is due to the mapping relationship between phase and far-field is more complex when the number of elements increases. If researchers want to obtain the same precision as the 7 elements system, they need to increase the number of training data.

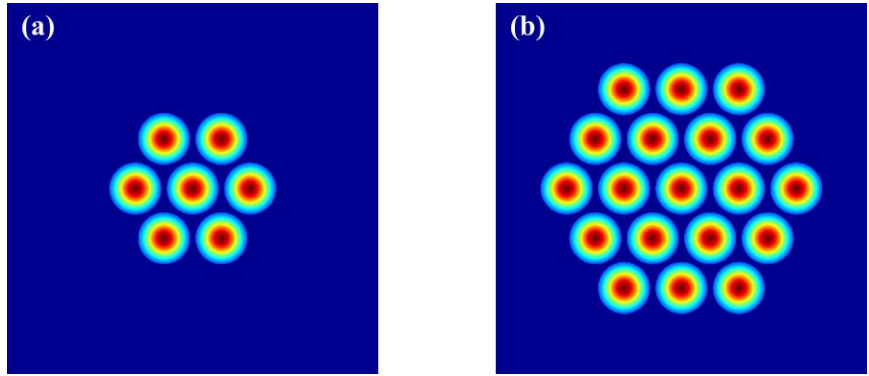

Fig. 15. Sketch of the emissive plane in the FLPA system. (a) 7 elements and (b) 19 elements

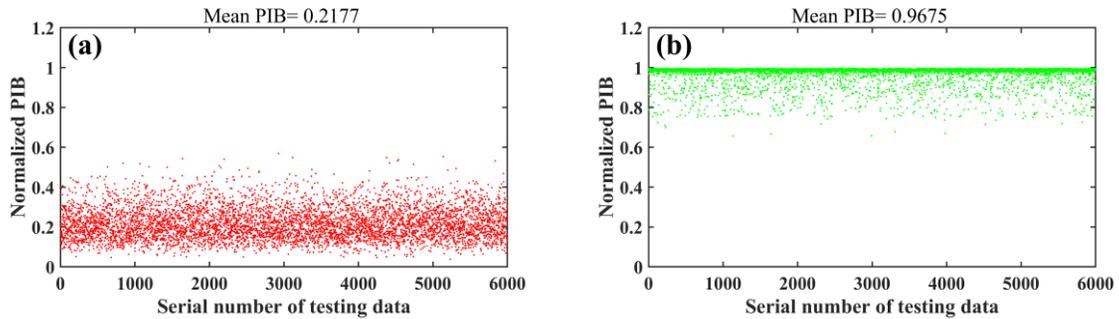

Fig. 16. PIB scatter plot of 19 elements FLPA system

(a) Initial PIB distributions, (b) PIB distributions after being compensated by CNN

## 5. Conclusion

We analyzed the requirement of high-precision phase-locking in the FLPA system. By using a pair of focal-plane images to reverse the initial piston aberration, the system could be directly entered into the convergence state. A method based on spiral phase shape modulation to break the phase ambiguity was proposed to solve the non-uniqueness issue, and a 1-1 mapping relationship between the initial piston aberration and the far-field images was established. To reduce the restoration time of piston aberration, CNN was introduced to overcome the multiple disturbance optimization processes of the traditional algorithm. In addition, a new evaluation index (NPCD) was proposed for a more proper CNN evaluation. The results showed that our proposed method could achieve high-speed and high-precision piston retrieval.

The PIB of 0.993 was simulated achieved. In our phase-locking strategy, the SPGD algorithm was no longer used to optimize the initial piston aberration, but to keep a co-phase state of the system. The system could directly enter the convergence state after initial compensation by CNN. This proposed method also meets the high-precision and robustness phase-locking requirements of large-scale FLPAs in a complex environment.

## Funding

## Disclosures

†Jing Zuo and Haolong Jia are co-first authors. The authors declare no conflicts of interest.